\documentstyle[aps,pre,twocolumn,graphicx,tabularx]{revtex}
\begin{document}
\draft
%
\wideabs{
%
\title{Semi-Analytical Calculation of the Rouse Dynamics of Randomly
Branched Polymers}
\author{Josh P. Kemp, Zheng Yu Chen}
\address{Guelph-Waterloo Program for Graduate Work in Physics and
Department of Physics, University of Waterloo, Waterloo, Ontario, Canada
N2L 3G1} 
\date{\today}
\maketitle
\begin{abstract}
We present a semi-analytical approach to the determination of the dynamic
properties of randomly branched polymers under the Rouse approximation. The
principle procedure is based on examining a spectrum of eigenvalues which
represents the average dynamic behavior of various structures. The calculated
spectra show that the eigenvalue distribution is random even within a single
structure which in turn produces a continuous spectrum of values for the
entire class. The auto-correlation functions for the radius of gyration
squared were calculated based on these spectra, which confirms that the
dynamics is non-exponential as earlier reported. A universal stretched
exponent is also found in this study. 
\end{abstract}
}

\begin{figure}
\includegraphics[width=8cm]{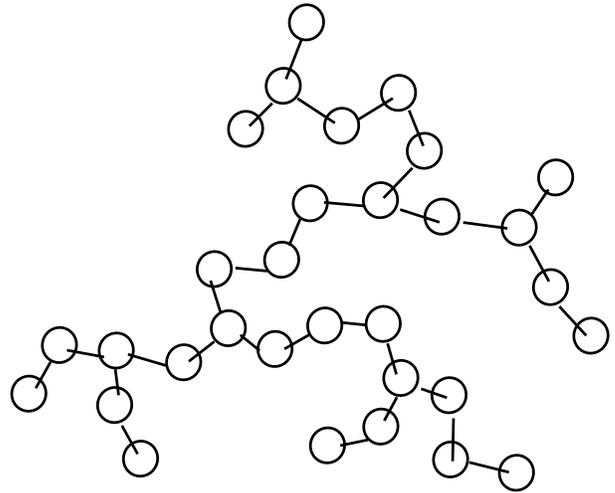}
\caption{Sketch of a typical randomly branched polymer.}
\label{rbp}
\end{figure}

	Among macromolecular structures found in many important systems such
as plastics, proteins, and sol-gel networks\cite{daoud,gennes,patton},
randomly branched structures occupy a unique place, as they display distinct
physical properties from those of linear or regularly branched polymers. A
typical molecule is constructed from a connection of linear polymer portions
of various lengths at branching sites selected randomly according to the
underlying physical branching mechanism [see Fig.  \ref{rbp}].  The study of
randomly branched polymers (RBPs) poses an interesting challenge to condensed
matter theorists, as the average over many different branching structures
must be included in addition to the ensemble average taken over the
configurational space for a given single structure.  Most of the previous
studies on these polymers have been focused on their conformational
properties\cite{parisi,cui,gutin}, which reveal different scaling properties
from linear polymer chains. The more intriguing aspect of randomly branched
polymers is probably their connection to the spin glass problem: there exists
a difference in quenched and annealed disorders, typical to random
systems\cite{cui,gutin}.  The dynamic properties of these polymers, however,
remains largely unexplored. 

	In a recent study\cite{kemp}, we were able to show, with the aid of
Monte Carlo computer simulations, that the dynamic behavior of these polymers
in a good solvent does not exhibit the conventional exponential relaxation
normally seen in other regularly branched structures.  The difficulty
associated with the dynamic study of this class of polymer is that each
polymer has a unique dynamic behavior dependent on its structure. Although
the structural averages can be performed using various techniques developed
in statistical physics\cite{gutin}, it is unclear how to construct a rigorous
analytical theory to explain the dynamic properties even for the simplest
cases, which have been examined by using scaling arguments\cite{daoud,kemp}.
In principle, numerical simulations of polymers provide an alternative
approach for examining the basic dynamic behavior of these randomly branched
structures, allowing the sampling of many different branching structures. In 
\noindent
practice, however, the number of branching structures that can be sampled is
limited, due to the extensive computational time required.

There are many unaddressed questions regarding the dynamics of RBPs. For
example, one simple question is whether or not the non-exponential behavior
is indeed an intrinsic property of the randomness. If it is, then it will
show up inherently even in the simplest type of polymer dynamics, the Rouse
dynamics, where the excluded volume and hydrodynamic interactions are both
neglected. In this article, we examine the Rouse dynamics of an
autocorrelation function and intrinsic viscosity of RBPs using a
semi-analytical method with a satisfactory average over the structural space. 

For a given structure, the Rouse dynamics are characterized by a $NxN$ Rouse
force matrix $A$ appearing in the Langevin equation\cite{doi},
\begin{equation}
\label{langevin}
\zeta{{d{\bf\vec r}}\over{dt}}= kA{\bf\vec r} + {\bf\vec f}(t)
\end{equation}
where $\zeta$ is the inverse mobility, ${\vec r}$ is a N-dimensional vector
containing the positional coordinates of the N 
\begin{figure}
\includegraphics[width=8cm]{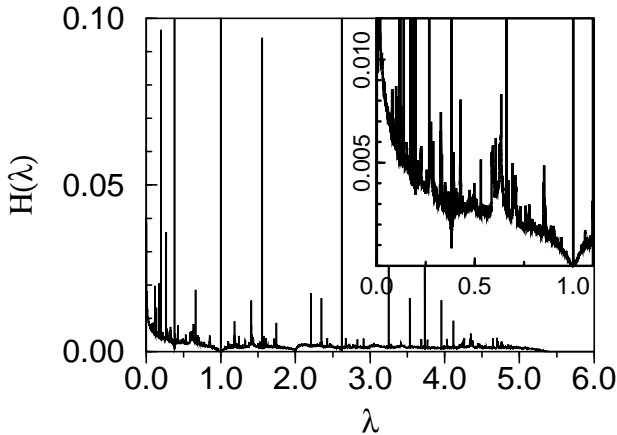}
\vspace{2mm}
\caption{The average eigenvalue spectrum for RBPs of
$N=100$. The spectrum appears continuous suggesting that the eigenvalues are
distributed randomly for a single structure. The inset graph shows
dominating eigenvalues in the long time scale, and
emphasizes the continuous distribution of the modes. The five peaks which
extend beyond the range of the main graph, reach values of
approximately 0.7, 2.5, 0.7, 0.1, and 0.03, from left to
right.} 
\label{hist101} 
\end{figure}
\noindent
monomers, and ${\vec f}(t)$ is
a $N$-dimensional random force vector containing Brownian forces acting on
these $N$ monomers. The $N$ eigenvalues of $A$ determine the relaxation rate
and dynamic properties of various collective normal modes of the given
structure. 

The matrix $A$ is directly determined by how the structure (Fig. \ref{rbp}) 
is labeled and connected, which is analytically generated here by a cut and
paste algorithm, similar to the one used by Cui and Chen\cite{cui}.  Starting
from a linear polymer, we choose a bond randomly and cut the molecule into
two portions. The smaller portion was then attached to a randomly chosen
monomer on the larger portion. The move was discarded if the anticipated
bonding site was already a branching monomer. The spatial positions of the
monomers were not kept track of as the Rouse matrix is only concerned with
the connectivity of the structure when the monomers have no volume. The
process was initially repeated $10^5$ times to create an equilibrate
structure before data collection. This algorithm was originally suggested by
Rensberg and Madras\cite{rensburg}, in their study of lattice ``animals", and
has been shown to yield a branching ratio between the number of branching
nodes, $n_3$, and total number of bonds, $N$, of
$n_3/N\approx0.25$\cite{cui}.

A Rouse matrix was created after every ten cut-paste moves, to ensure that
the structures were not all correlated. Once the Rouse matrix was defined, a
complete set of eigenvalues were then determined numerically\cite{numrec},
and binned to create a normalized histogram.  Each bin was given a width of
$1.0\times 10^{-4}$ to create a fine scale image of the distribution.  For
each given $N$, a total of $10^{7}$ structures were examined.

\begin{figure}
\includegraphics[width=8.5cm]{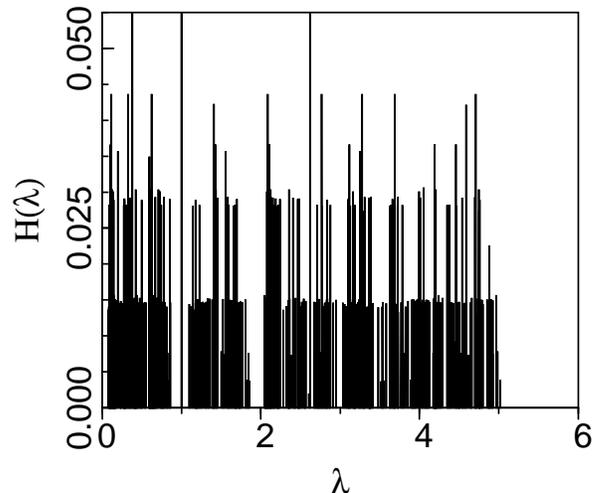}
\caption{The average eigenvalue spectrum for RBPs of
$N=10$. The small size of the polymer generates a distinctly different
spectrum from that in Fig. \ref{hist101}. The spectrum is discrete,
yet
randomly distributed. This spectrum still
generates a non-exponential behavior.} 
\label{hist11} 
\end{figure}

The spectrum that is produced is distinctly different from that of a linear
polymer. The spectrum for a linear polymer would be constructed of only $N$
eigenvalues, and thus would appear as $N$ discrete lines spaced according to
the following equation\cite{doi},
\begin{equation}
\lambda_p=\lambda_1\ p^2\ ;\ p=1,2,\dots,N\ ,
\end{equation}
where $\lambda_1$ is the smallest eigenvalue. In Fig. \ref{hist101} the
eigenvalue histogram for a RBP structure of $N=100$ bonds is displayed. The
magnified portion for eigenvalues between 0 and 1, which governs the longest
time scales of relaxation, shows that the distribution is continuous, and
roughly follows a power law near $\lambda=0$. This histogram distribution
indicates that the eigenvalues are uniformly distributed due to various
segmental lengths and the coupling of the segments in the molecules.  The
figure also demonstrates that the eigenvalues of a single structure are
randomly distributed, opposed to an ordered distribution as in a linear
polymer. These features can be seen more clearly in Fig. \ref{hist11}, where
the histogram for N=10 is displayed. In a structure this small there are a
limited number of distinct structures, and thus a limited number of possible
eigenvalues, so the eigenvalues appear discrete in the spectrum. The figure
clearly shows how the eigenvalues are randomly arranged in the spectrum. 

Returning to Fig. \ref{hist101}, there appear to be several eigenvalues which
occur more frequently than the average. These eigenvalues are most likely
caused by frequently occurring sub-structures that are regularly branched or
linear. For example, the relative motion of a sub-unit that has two outer
monomers connected to a stationary branching unit, will have an eigenvalue of
1\cite{cai}; the large peak indicates the frequent appearance of such units. 

One of the most direct probes of the internal structure is the
autocorrelation function associated with the radius of gyration squared. As
has been appreciated in other 
\begin{figure}
\includegraphics[width=8cm]{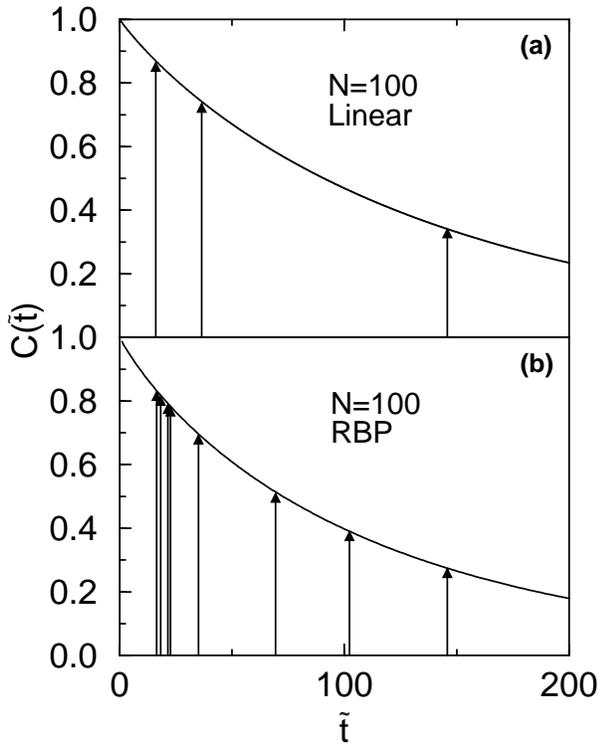}
\caption{Auto-correlation functions for a linear polymer (a) and a RBP
(b) of length N=100.
The function in (a) has been scaled such that the largest
characteristic time, $\mathaccent126{\tau_1}$,
is set to match that in figure (b). The other largest characteristic times  
are also
placed on the graph to demonstrate their magnitudes. In figure (b),
long time relaxation is dominated by multiple modes, which cause the
observed non-exponential behavior.} 
\label{scale} 
\end{figure}
\noindent
macromolecular structures, such a correlation
function
characterizes the radial relaxation motion, and defines the so-called elastic
relaxation for a macromolecule\cite{murat}. The dynamic relaxation of such a
correlation is comprised of the various internal relaxation times. Due to the
large number of small eigenvalues in a RBP, no eigenvalue can be singled out
to yield dominating exponential behavior.  More specifically, we examine
\begin{eqnarray}
\label{cs2}
C(t) & = & {{<S^2(t)\cdot S^2(0)>-<S^2(0)>^2}\over {<S^2(0)\cdot
S^2(0)>-<S^2(0)>^2}}\nonumber \\
     & = &{{\sum_{i=1}^{N}{H(\lambda_i)\over\lambda_{i}^{2}}e^{-\lambda
_{i}{\mathaccent126{t}}}}\over
{\sum_{i=1}^{N}{H(\lambda_i)\over\lambda_{i}^{2}}}}
\end{eqnarray}
where $\mathaccent126{t}$ is the rescaled time such that
$\mathaccent126{t}={2kt/\zeta}$, and $H(\lambda_i)$, represents the
probability of the occurrence of the eigenvalue $\lambda_i$ given by the
histogram.

	To further illustrate the significance of many time scales involved
in the dynamics, the autocorrelation function for the radius of gyration
squared (Eq.\ref{cs2}) was calculated for a linear polymer and a typical RBP
of N=100 monomers in Fig. \ref{scale}a and Fig. \ref{scale}b respectively.
Also plotted in the figures are the longest characteristic time 
\begin{figure}
\includegraphics[width=8cm]{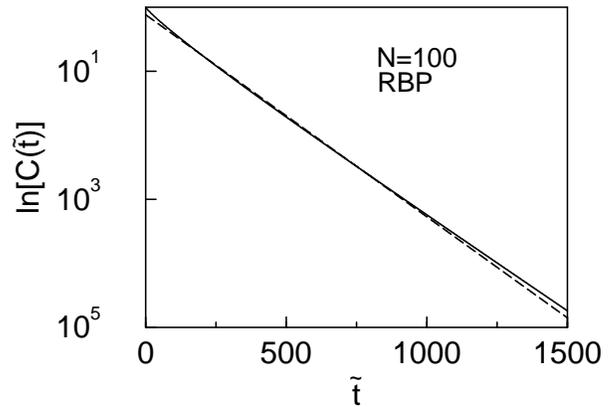}
\caption{A semi-log plot of the correlation function vs. rescaled time, for
the RBP in figure \ref{scale}b. The curve is fitted with an exponential
function to demonstrate the systematic deviation of the curve from this
function,
implying a non-exponential relaxational behavior.} 
\label{fit}
\end{figure}
\noindent
scales,
indicated by the vertical arrows. The RBP has a significantly shorter
relaxation time than the linear polymer since the structure is more compact,
thus in the comparison of the distribution of the characteristic times, the
linear polymer's time scales have been scaled such that the longest
characteristic times of both polymers match exactly. Of course, when $t$ is
extremely large only the largest time scale is dominant, but this only occurs
when the correlation is no longer significant, thus the correlation function
is dominated by non-exponential relaxation. Comparing the two figures, we see
that the characteristic time scales in a linear polymer, in which
$\tau_p=\tau_1p^{-2}, p=1,2,\dots,N$\cite{doi}, are sparsely spaced in the
region of interest, while the time scales in the RBP are irregularly
distributed in the same regime. The two or three longest time scales in Fig.
\ref{scale}b, are sufficiently close together to compete with each other; the
net result is a multi-exponential function rather than a single exponential
function as in Fig. \ref{scale}a. When averaged over different structures,
various long time scales must be taken 
\begin{figure}
\includegraphics[width=8cm]{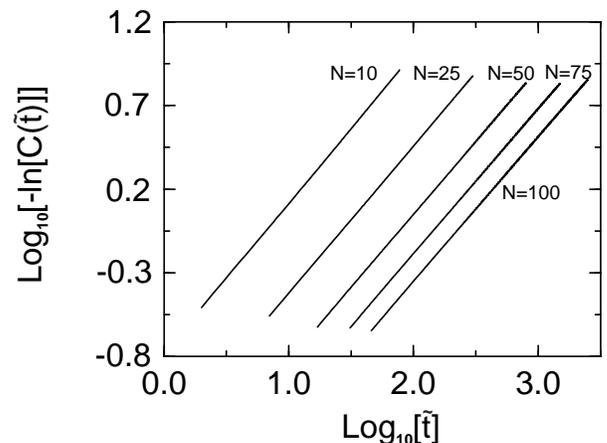}
\caption{A double logarithmic plot of the correlation functions, for
N=10,25,50,75,and 100. These plots are used to calculate the parameters
$\alpha$ and $\mathaccent126{\tau}$ in Eq. (\ref{stretched}).} 
\label{dlog}
\end{figure}
\noindent
into account, which yields an average
over various multi-exponential functions. As discussed by Palmer et
al.\cite{palmer}, and Iori et al.\cite{iroi}, such a process produces much
slower decaying correlation functions, which sometimes are represented by
various non-exponential functions. 

	The co-operative behavior of the competing time scales can be seen in
Fig. \ref{fit}, from another perspective. The figure shows a semi-log plot of
autocorrelation function for the same function in Fig. \ref{scale}b.  The
dashed line is the best fit to the solid curve with an exponential function,
which demonstrates that the correlation function is not exponential as in
linear polymers. This result is in agreement with previous Monte Carlo
simulation results which show the same non-exponential relaxation\cite{kemp}. 

	We have shown thus far that the eigenvalues of RBPs are randomly
distributed, and the continuous distribution near $\lambda=0$ yields insight
into the non-exponential relaxation. Without further knowledge of an
analytical expression of $H(\lambda)$, the summation in Eq.(\ref{cs2}) 
cannot be carried out exactly.  As in previous work\cite{kemp}, due to
similarities of this system with spin glass systems and other random
disordered systems\cite{palmer,iroi,nixon,ogielski}, it is instructive to
compare the correlation curve to a stretched exponential.
\begin{equation}
\label{stretched}
C_{st}(\mathaccent126{t})=e^{[-\mathaccent126{t}/\mathaccent126{\tau}
]^{\alpha}}
\end{equation}
where $\mathaccent126{\tau}$ represents a general characteristic time for the
entire class of RBPs, and $\alpha$ is the stretched exponent. As can be
inferred from Fig. \ref{dlog}, the stretched exponential function gives a
rather good approximation with the curves produced from Eq. (\ref{cs2}). The
$\alpha$ values for different $N$ are shown in Table \ref{results}, which
demonstrates that $\alpha\approx0.87\pm0.01$ appears universal for all large
$N$ considered here. The range used to calculate the exponents was selected
by choosing the range of time scales such that the auto-correlation function
was approximately $0.8$ to less than $10^{-4}$, corresponding to the region
where an exponent of one can be produced for linear polymers.  This value can
be compared with $\alpha\approx0.8$ exponent calculated in earlier
work\cite{kemp}; the discrepancy is not surprising since this model considers
no excluded volume effects.

To further determine the universality of the stretching exponent $\alpha$, we
have calculated the eigenvalue spectra for RBPs with a different branching
probability, corresponding to $n_3/N= 1/2$. In these polymers there exist
only end and branching points in the structure, and no linear 
\begin{table}
\caption{Stretched exponents, relaxation times and intrinsic viscosity
determined for various $N$. These systems correspond to a branching
ratio of $n_3/N\approx0.25$.} 
\label{results}
\begin{center}
\begin{tabular}{@{\hspace{0.25cm}}c@{\hspace{0.5cm}}c@{\hspace{0.5 
cm}}c@{\hspace{0.5cm}}c@{\hspace{0.25cm}}}
\hline\hline
{\bf $N$} & {\bf $\alpha$}   & {\bf $\mathaccent126{\tau}$}	&{\bf
$[\mathaccent126{\eta}]$} \\ \hline
$10$	  & $0.902\pm0.005$  & $7.55\pm0.05$	& $1.49\pm0.05$  \\
$25$	  & $0.879\pm0.005$  & $30.2\pm0.05$	& $2.81\pm0.05$  \\
$50$	  & $0.871\pm0.005$  & $87.8\pm0.05$	& $4.38\pm0.05$  \\
$75$	  & $0.865\pm0.005$  & $163.6\pm0.05$	& $5.63\pm0.05$  \\
$100$	  & $0.861\pm0.005$  & $254.2\pm0.05$	& $6.69\pm0.05$  \\
\hline\hline
\end{tabular}
\end{center}
\end{table}
\begin{figure}
\includegraphics[width=8cm]{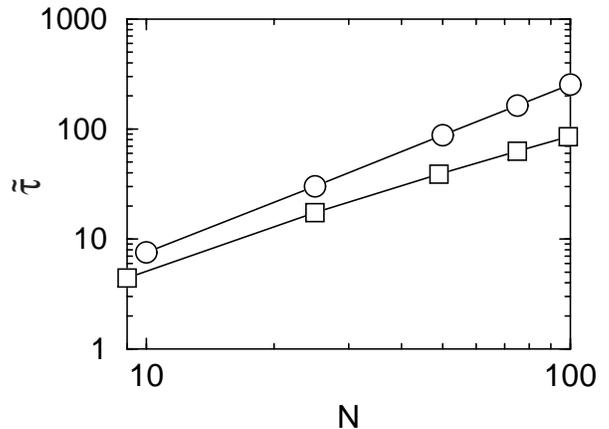}
\caption{A log-log plot of $\mathaccent126{\tau}$ vs. $N$. 
Circles and
squares correspond to two different classes of RBPs, those with
$n_3/N=0.25$ and $n_3/N=0.5$ respectively.} 
\label{tau}
\end{figure}
\noindent
parts are
allowed. To create these structures, a base star nucleus of 4 monomers was
used. Using this base an end 
monomer was randomly selected and two monomers
were attached to change the end monomer to 
a branching point. This process
was repeated until the desired structure size, $N$ = 9, 25, 49, 75, and 99,
was reached. The resulting correlation curves demonstrate the same stretched
exponent as in the case of much lower branching probability. The results are
shown in Table \ref{result2}.  This suggests that there is little or no
dependence of the stretching exponent on the branching probability. Also, the
examination of the exponents suggests that there is an insignificant
correlation with $N$, suggesting that this stretching behavior is universal
to the entire class of RBPs. Whether the quenched and annealed universality
classes\cite{cui,gutin} will produce the same exponents cannot be determined
without the introduction of the excluded volume. 

The exponent $\mathaccent126{\tau}$, displayed in Tables \ref{results} and
\ref{result2}, in equation \ref{stretched} can also be calculated from Fig.
\ref{dlog}. The plot against $N$ in Fig.\ref{tau} yields the scaling laws,
\begin{eqnarray}
\label{tscaleeq}
\mathaccent126{\tau} &\sim&  N^{(1.5\pm0.1)}\ for\  n_3/N\approx0.25 \\
\mathaccent126{\tau} &\sim&  N^{(1.2\pm0.1)}\ for\  n_3/N\approx0.5 
\end{eqnarray}
Our early work\cite{kemp} suggests a larger exponent with the inclusion of an
excluded volume interaction. This is consistent with the fact that the
excluded volume effect slows 
down the dynamics generally\cite{downey}. 
\begin{table}
\caption{Stretched exponents, relaxation times, and intrinsic viscosity
for various $N$. These systems correspond to a branching ratio of
$n_3/N\approx0.5$.} 
\label{result2}
\begin{center}
\begin{tabular}{@{\hspace{0.25cm}}c@{\hspace{0.5cm}} 
c@{\hspace{0.5cm}}c@{\hspace{0.5cm}}c@{\hspace
{0.25cm}}}\hline\hline
{\bf $N$}   & {\bf $\alpha$}	& {\bf $\mathaccent126{\tau}$}	& {\bf
$[\mathaccent126{\eta}]$} \\ \hline
$9$	    & $0.924\pm0.005$	& $4.4\pm0.05$	& $1.20\pm0.05$  \\
$25$	    & $0.874\pm0.005$	& $17.3\pm0.05$	& $2.24\pm0.05$  \\
$49$	    & $0.859\pm0.005$	& $39.0\pm0.05$	& $3.15\pm0.05$	 \\
$75$	    & $0.851\pm0.005$	& $63.0\pm0.05$	& $3.80\pm0.05$	 \\
$99$	    & $0.846\pm0.005$	& $85.1\pm0.05$	& $4.25\pm0.05$	 \\
\hline\hline
\end{tabular}
\end{center}
\end{table}
\begin{figure}
\includegraphics[width=8cm]{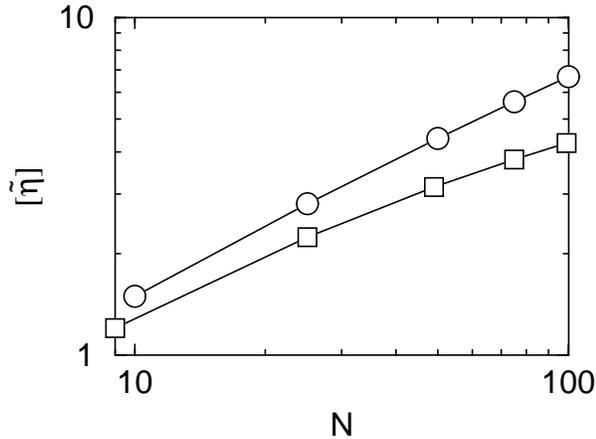}
\caption{A log-log plot of the intrinsic viscosity vs. $N$. 
Circles and
squares correspond to two different classes of RBPs, those with
$n_3/N=0.25$ and $n_3/N=0.5$ respectively. The straight line represents
Eq.(\ref{viscosity}).} 
\label{visc}
\end{figure}

The intrinsic viscosity, $[\eta]$, is also a significant probe of the
internal dynamics of a macromolecule, and can be directly measured
experimentally. One can show that $[\eta]$ is related to the
eigenvalues\cite{zimm} by
\begin{equation}
\label{viscosity}
[\mathaccent126{\eta}]={{[\eta]6\rho\eta_s}\over {N_Aa^2\zeta}} 
= {1\over
{(N+1)}}\sum_{i=1}^{N}{{H(\lambda_1)}\over\lambda_{i}}
\end{equation}

where $N_A$ is Avogadro's number, $\eta_s$ is the viscosity of the solvent,
$\rho$ is the weight per monomer, and $a$ is the intermolecular bonding
distance.  The rescaled intrinsic viscosities are displayed in Tables
\ref{results} and \ref{result2}.  The log-log plot of
$[\mathaccent126{\eta}]$ vs. $N$ shows
\begin{eqnarray}
\label{vscaleeq}
[\mathaccent126{\eta}] &\sim& N^{0.63\pm0.01}\ for\ n_3/N\approx0.25 \\
\lbrack\mathaccent126{\eta}\rbrack &\sim& N^{0.47\pm0.01}\
for\ n_3/N\approx0.5
\end{eqnarray}
The exponent calculated here is for a Rouse model dynamics.  It is known that
under the Rouse approximation, no simple scaling relation between [$\eta$]
and $N$ can be deduced. While linear polymers show $\eta\propto N$,
star-burst dendrimers show $ \eta\propto ln N$\cite{cai}.

In practical good solvents, both hydrodynamic and excluded-volume effects
must be considered. In such cases, a simple relationship can be deduced based
on the fact that [$\eta$] is inversely proportional to the particle density
$S^3 /N$, where $S$ is the radius of gyration. Based on this scaling
argument, Daoud et al.\cite{daoud} have concluded a scaling exponent of $3/8$
for [$\eta$], which, of course, cannot be directly compared with the exponent
found here due to different physical origins involved.  The exponent $3/8$
agrees with the experimental results of branched polyethylene by Patton et
al.\cite{patton}. A more rigorous treatment of [$\eta$] for RBPs beyond the
scaling argument, however, cannot be found in previous literatures.

	In summary, by examining the average eigenvalue spectrum of RBPs, we
were able to show that the eigenvalues are randomly distributed as opposed to
periodically organized. The result of this distribution is the competition of
relaxational modes which leads to a non-exponential behavior in the
relaxation of the polymer. Fitting the auto-correlation functions of the
radius of gyration squared to a stretched exponential we found a stretching
exponent of $\alpha = 0.87 \pm 0.04$.

We would like to thank C. Cai for providing the algorithm to calculate the
eigenvalues, and S. Dwyer for the critical reading of the manuscript. We
would also like to thank the Natural Sciences and Engineering Research
Council of Canada for financial support. 


%
%
%
%

%
%

\end{document}